**Research article**

# A Dual-Dip Heterogeneous LPFG Sensing System via Annealing under Bending with Temperature and Humidity Compensation

Cuiying Huang[1], Riming Xu[2], Jialing Kang[3], Weihan Chen[3], Xingnan Chen[4], Yanbo Li[5], Jin Wang[2, 6]*

1. School of Arts and Sciences, Fujian Medical University, Fuzhou 350122, China.
2. Physical Sciences and Engineering Division, King Abdullah University of Science and Technology, Thuwal 23955-6900, Saudi Arabia.
3. School of Basic Medical Sciences, Fujian Medical University, Fuzhou 350122, China.
4. Collage of Mechanical and Electrical Engineering, Hainan Vocational University of Science and Technology, Haikou 571125, China.
5. Department of Civil and System Engineering, Johns Hopkins University, Baltimore 21218, USA.
6. College of Physics and Optoelectronics, Faculty of Science, Beijing University of Technology, Beijing 100124, China.

*Corresponding author: Jin Wang
E-mail: wangjinbjut321@163.com

**Abstract**

Optical fiber multi-parameter sensing is fundamentally constrained by cross-sensitivity and the complexity of multi-sensor integration. Here, we present a dual-dip heterogeneous long-period fiber grating (LPFG) sensing platform enabled by bending-assisted annealing, which introduces anisotropic refractive index redistribution and mode-dependent coupling enhancement. This process yields enhanced sensitivity, improved dip contrast, and opposite spectral responses between dual resonance dips, providing intrinsic spectral heterogeneity. To overcome temperature cross-sensitivity, a polymer-encapsulated cascaded LPFG–FBG architecture is developed, where the LPFG serves as the microbending-sensitive element and the FBG acts as a reference channel. PDMS encapsulation enhances stress transfer and suppresses interfacial slippage, improving linearity and repeatability. As a result, the bending sensitivity increases from −3.44 to −8.97 nm/cm, and the detection limit improves from 0.017 to 0.006 cm. Building on this, a multi-parameter sensing paradigm is established by integrating dual-dip heterogeneity with LPFG–FBG spectral orthogonality. With PAAm functionalization, the platform enables simultaneous and decoupled sensing of temperature, bending, and humidity, demonstrating scalable and versatile multi-parameter capability. Overall, this work establishes a minimalistic yet robust paradigm for multi-parameter fiber-optic sensing, offering a scalable strategy for high-performance sensing in structural health monitoring and harsh environments.

**Introduction**

Optical fiber sensors have attracted extensive attention for structural health monitoring owing to their advantages of high sensitivity, immunity to electromagnetic interference, and capability for distributed sensing [1-3]. In particular, long-period fiber grating (LPFG) and fiber Bragg grating (FBG) have been widely employed for the detection of physical parameters such as temperature, strain, bending, and other environmental variations [4-5]. However, in practical applications, especially in complex environments such as pipeline monitoring, the simultaneous measurement of multiple parameters remains a significant challenge [6-7]. A fundamental limitation of optical fiber sensing lies in cross-sensitivity, where multiple variables perturbations induce overlapping spectral responses.

Traditional approaches to address this issue typically rely on combining multiple sensors or introducing additional reference elements [8-9]. Although effective, these strategies often lead to increased system complexity, higher fabrication cost, and more demanding signal demodulation procedures. Moreover, when identical grating sensors are used, the intrinsic homogeneity of their sensing characteristics results in similar responses to different perturbations, making it difficult to construct a reliable multi-parameter sensing matrix. To overcome these challenges, recent efforts have focused on introducing spectral diversity or structural asymmetry within the sensing system. However, achieving sufficient heterogeneity typically requires complex system designs or multiple sensing elements, which limits scalability and robustness [10-11].

In this work, we propose a fundamentally different strategy by exploiting intrinsic heterogeneity within a single LPFG, achieved through a bending-assisted annealing process.

This approach induces anisotropic refractive index redistribution and mode-dependent coupling enhancement, leading to dual resonance dips with opposite spectral shift trends and distinct sensitivities. Such built-in heterogeneity provides a natural foundation for multi-parameter sensing without increasing system complexity.

To further address temperature cross-sensitivity, a polymer-encapsulated cascaded LPFG–FBG architecture is developed. In this configuration, the LPFG serves as a highly sensitive bending sensor, while the FBG functions as a temperature reference channel. The incorporation of a viscoelastic polydimethylsiloxane (PDMS) layer enhances stress transfer efficiency and improves sensing stability, while maintaining high linearity and repeatability. In addition, by introducing functional polymer coatings such as polyacrylamide (PAAm), the sensing system is extended to include humidity responsiveness. By combining spectral heterogeneity, structural decoupling, and functional material integration, a multi-wavelength demodulation matrix is established, enabling simultaneous and decoupled measurement of temperature, bending stress, and humidity.

This work provides a minimalistic yet powerful paradigm for multi-parameter fiber-optic sensing, offering significant potential for applications in pipeline monitoring, wearable sensing, and harsh-environment detection.

## Annealing under bending conditions to enhance dip sensitivity and heterogeneity

In this work, as illustrated in **Fig. 1a**, the LPFG was annealed under a controlled bending condition. All details of the materials and measurement procedures are provided in **Supplementary Note 1**. A bend annealing enhanced mode coupling mechanism was established to enhance the sensitivity and heterogeneity of the characteristic resonance dips of the sensor.

LPFG operated based on phase matching between the fundamental core mode and discrete cladding modes [12-13]. The resonance wavelength of the m-th cladding mode can be expressed as,

$$\lambda_{res,m} = \left(n_{co,eff} - n_{cl,eff,m}\right)\ \Lambda \tag{1}$$

where $n_{co,eff}$ and $n_{cl,eff,m}$ denote the effective refractive indices of the core and the m-th cladding modes, respectively, and $\Lambda$ is the grating period.

During bend annealing, the LPFG is subjected simultaneously to a controlled curvature and elevated temperature [14-15]. This process introduces a coupled thermo-mechanical interaction involving (i) bending-induced stress distribution, (ii) stress relaxation dynamics of silica glass, and (iii) mode-dependent refractive index evolution. Bending introduces a transverse stress gradient, while annealing induces structural relaxation of the silica network, jointly modifying the effective refractive indices and mode coupling characteristics. Specifically,

For a fiber bent with radius $R$ (curvature $\kappa = 1/R$), the axial strain across the cross-section can be approximated as,

$$\varepsilon(x) = \kappa x \tag{2}$$

where $x$ is the transverse coordinate relative to the neutral axis, the $\kappa$ is the coupling coefficient. According to the photoelastic effect, the refractive index perturbation induced by bending is,

$$\Delta n(x) \approx -\frac{n^3}{2} p_e \kappa x \tag{3}$$

where $p_e$ is the effective photoelastic coefficient. This expression indicates that bending introduces a transverse refractive index gradient, resulting in asymmetrical modal redistribution.

When the bent LPFG is heated, residual stresses originating from fiber drawing and $CO_2$ inscription undergo thermally activated relaxation [16]. The structural rearrangement of the silica network follows viscoelastic relaxation kinetics, which can be described using a stretched-exponential function:

$$S(T,t) = 1 - exp\left[-\left(\frac{t}{\tau(T)}\right)^{\beta}\right] \tag{4}$$

where $S(T,t)$ represents the relaxation progress, $\beta$ is the stretching exponent ($0 < \beta \leq 1$), and the characteristic relaxation time follows an Arrhenius relation,

$$\tau(T) = \tau_0 \ exp\left(\frac{E_\alpha}{k_B T}\right) \tag{5}$$

Here, $E_a$ is the activation energy and $k_B$ is Boltzmann's constant.

Under bending conditions, stress relaxation becomes spatially asymmetric. Part of the bending-induced refractive index perturbation is "frozen" after cooling, forming a quasi-permanent anisotropic refractive index distribution. Consequently, the total refractive index perturbation can be expressed as the sum of reversible and permanent components,

$$\Delta n(x;\kappa,T,t) = \Delta n_{rev}(x;\kappa) + \alpha n_{rev}(x;\kappa) \ S(T,t) \tag{6}$$

where $\alpha$ represents the fraction of bending-induced stress that becomes permanently retained after annealing.

Different cladding modes exhibit distinct spatial field distributions and boundary sensitivities, resulting in different responses to asymmetric refractive index perturbations. The resonance wavelength evolution can be approximated as,

$$\Delta\lambda_m(\kappa,T,t) = a_m\kappa + b_m\kappa S(T,t) + c_m S(T,t) \tag{7}$$

where $a_m$, $b_m$, and $c_m$are mode-dependent coefficients. Since the coefficients vary among modes and may differ in sign, characteristic dips may shift in opposite spectral directions as shown in **Fig. 1b**, thereby enhancing dip heterogeneity.

The coupling coefficient (κ) depends on refractive index modulation and modal overlap. Bend annealing improves modulation effectiveness and enhances modal overlap, leading to increased κ. At phase matching, the minimum transmission can be approximated by,

$$T_{min} \approx cos^2(\kappa L) \tag{8}$$

Thus, as $\kappa$ increases with annealing progress, the resonance dip deepens. The deepened dips provide improved signal-to-noise ratio and higher wavelength tracking precision, effectively enhancing the practical sensing resolution.

The post-annealing bending sensitivity becomes,

$$\frac{\partial\lambda_m}{\partial\kappa} = a_m + b_m\kappa S(T,t) \tag{9}$$

Since $S(T,t) \to 1$ with sufficient annealing time, the bending sensitivity increases [17-18]. As the curvature radius of the cylindrical tube to which the LPFG was attached decreased, the monitored spectral dip shifted toward longer wavelengths (**Fig. 1c**). Within the curvature radius range of 6–14.5 cm, the bending sensitivity of the dip varied from -2.21nm/cm to -3.47 nm/cm (**Fig. 1d**). Moreover, the difference in $b_m$ between distinct cladding modes amplifies differential spectral response, which is advantageous for multi-parameter decoupling and structural health monitoring applications. In summary, bend annealing introduces a controlled anisotropic refractive index redistribution and stress relaxation process that simultaneously enhances mode coupling strength, resonance contrast, bending sensitivity, and dip heterogeneity in LPFG sensors.

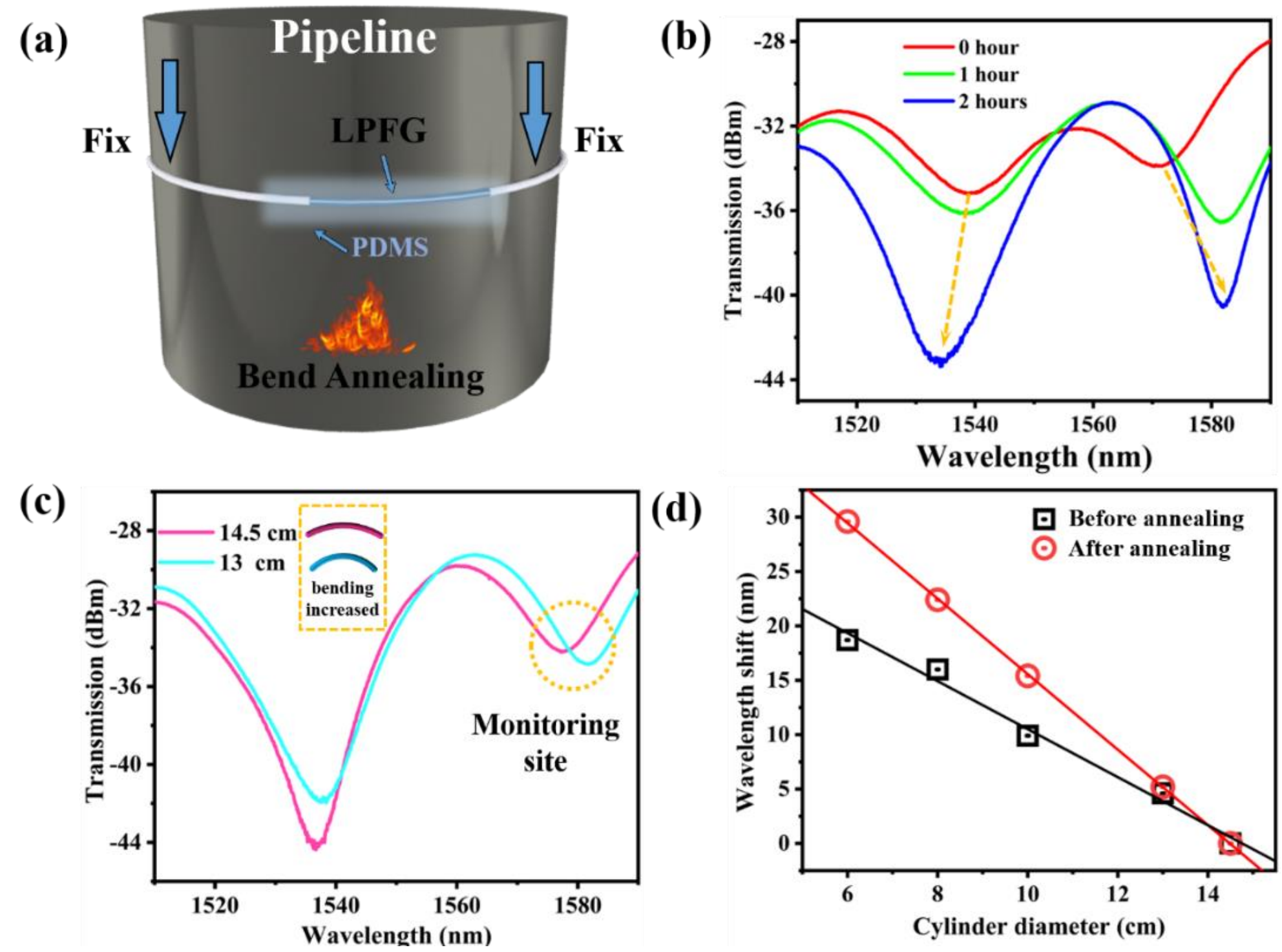

**Fig. 1** Enhancement of stress sensitivity and heterogeneity of an LPFG by annealing under a bending state. (a) Schematic of LPFG annealing under a bending condition. (b) Spectral evolution with increasing annealing time. As the annealing time increases, the coupling depth of the characteristic dip becomes more pronounced and shifts in the opposite spectral direction. (c) Spectral response of the LPFG to bending after annealing. The lateral shift of the monitored dip site was used to evaluate bending sensitivity. (d) Comparison of bending sensitivity before and after annealing. The bending sensitivity increased from -2.21nm/cm to -3.47 nm/cm.

**Practical bending monitoring of human joints and industrial pipelines**

The practical applicability of the proposed system for bending sensing was demonstrated through monitoring bending in human joints and strain in pipelines. The optical fiber was tightly attached to the wrist, as shown in **Fig. 2a–d**. As the wrist gradually changed from a fully extended state to a bending angle of 90°, the characteristic resonance dip exhibited a redshift toward longer wavelengths. The sensitivity corresponding to the variation of bending angle reached 7.1 nm/° based on the approximate statistics of the bending states.

The resonance dip also exhibited excellent reversibility during repeated bending and extension. At the same wrist bending position, the wavelength deviation of the dip was only 0.3–0.5 nm, demonstrating the capability to clearly distinguish different wrist bending states. This feature indicates strong potential for real-time monitoring in medical rehabilitation applications. It should be noted that this experiment primarily demonstrates the potential of the sensing system for wearable human-motion monitoring, while residual measurement errors may still arise from the normalization and calibration procedures.

In addition, the grating sensing system was attached to the sidewall of a pipeline, as illustrated in **Fig. 2f**. When the pipeline undergoes deformation under external pressure, the change in the curvature radius of the pipe induces bending of the grating simultaneously. By quantitatively compressing the pipeline, the sensing range was calibrated for curvature radii between 14 and 15 cm. As the curvature radius decreased, the characteristic resonance dip exhibited a redshift toward longer wavelengths. The sensitivity during the decreasing-curvature stage was −3.44 nm/cm, with a linearity coefficient of 0.98, while during the increasing-curvature stage the sensitivity was −3.38 nm/cm, with a linearity coefficient of 0.96 (**Fig. 2g**).

Owing to the nature of optical signal transmission, the spectral response occurs almost instantaneously with negligible hysteresis. Therefore, this sensing system is highly suitable for remote real-time monitoring of pipeline stress states, helping to prevent structural failure caused by excessive loading. However, slight deviations in repeatability and a modest reduction in linearity were observed during cyclic bending measurements. This effect mainly arises from minor misalignments of the fiber at its original attachment position caused by variations in the pipeline curvature radius. To address this issue, polymer encapsulation was subsequently introduced to stabilize the fiber position. Meanwhile, the polymer layer also enhances sensitivity and improves measurement stability, enabling more reliable sensing performance in practical applications. In addition, the inherent cross-sensitivity in fiber-optic sensing is an issue that cannot be neglected in practical measurements, particularly the interference induced by temperature variations. Therefore, effective compensation of temperature effects is also essential for achieving accurate and reliable sensing performance.

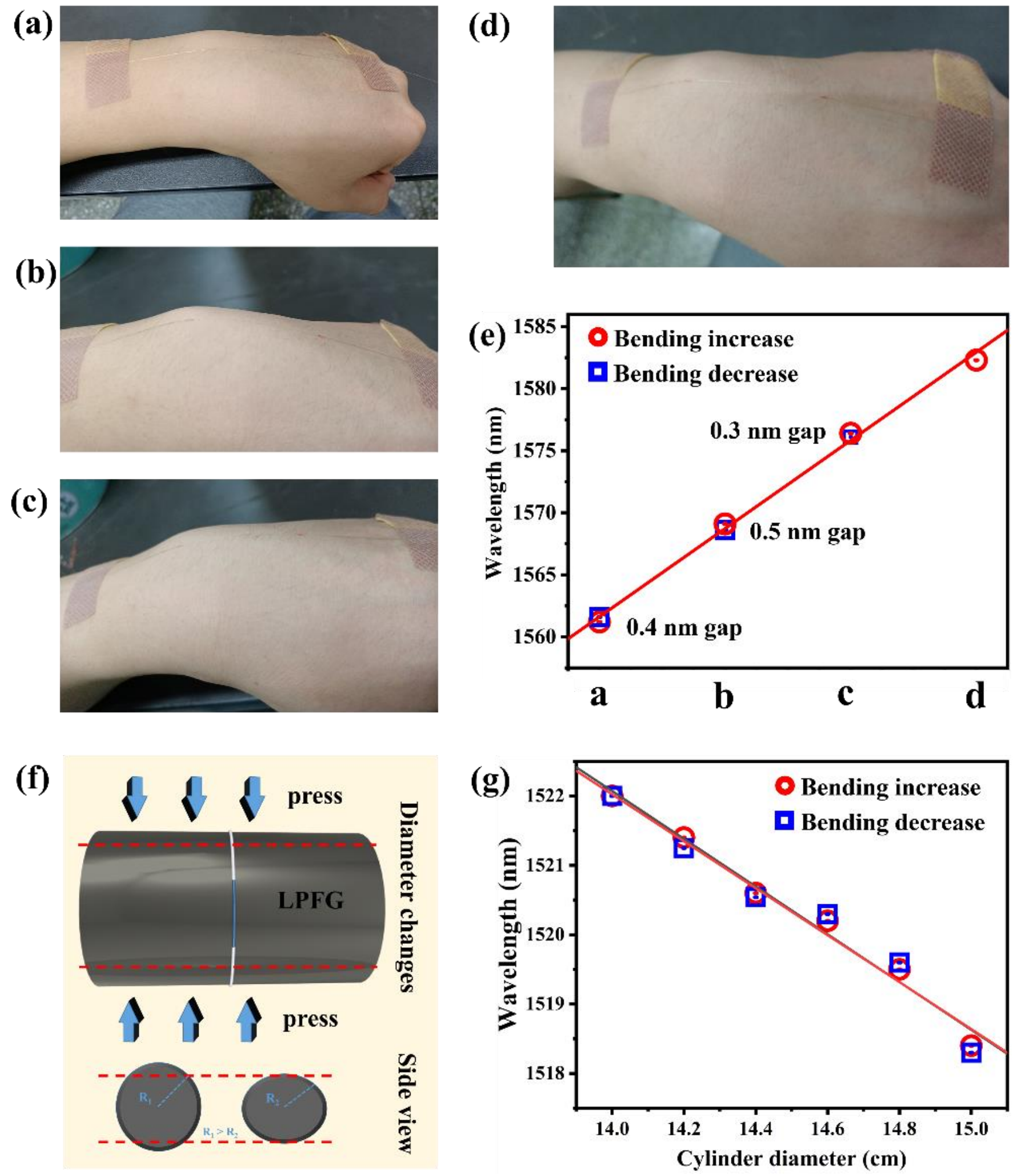

**Fig. 2** Bending monitoring of human joints and industrial pipelines. Demonstration of practical application in human wrist bending. LPFG attached to the wrist under different bending states: (a) fully extended; (b) bent to approximately 30°; (c) bent to approximately 60°; (d) bent to 90°. (e) Repeated wrist bending and extension test for evaluating the bending measurement stability of the LPFG. (f) Schematic diagram of pipeline compression monitoring. (g) Repeatability monitoring of curvature variation.

## Polymer-encapsulated cascaded system for temperature-compensated bending stress monitoring

Cross-sensitivity is an intrinsic limitation in fiber-optic sensing systems, arising from the overlapping responses of different physical parameters. To overcome this constraint, we develop a temperature-decoupled bending stress sensing platform based on a cascaded LPFG–FBG architecture, in which structural anisotropy and spectral orthogonality are simultaneously exploited. The detailed experimental setup is presented in **Supplementary Note 2**.

As shown in **Fig. 3a**, the LPFG and FBG are serially integrated to form a cascaded sensing unit with differentiated mechanical coupling pathways. The LPFG is conformably attached along the radial direction of the pipeline, where external pressure-induced deformation is efficiently converted into bending perturbations, thereby modulating the phase-matching condition and enhancing the sensitivity of the resonance dips. In contrast, the FBG is aligned along the axial direction, where bending-induced strain is effectively minimized, allowing it to function as a thermally responsive yet mechanically isolated reference channel (a stress-insensitive FBG is further employed to suppress residual strain cross-coupling), as illustrated in **Fig. 3b**. The spectral response of the cascaded system (**Fig. 3c**) exhibits well-separated modal features, where the LPFG resonance dips and the FBG reflection peak occupy distinct spectral domains. This spectral decoupling enables independent tracking of stress- and temperature-induced wavelength shifts. By constructing a multi-wavelength demodulation matrix, the coupled perturbations can be quantitatively resolved, thereby achieving simultaneous and decoupled measurement of external pressure and ambient temperature.

Furthermore, to enhance the sensing performance, polymer encapsulation was implemented in conjunction with the fiber annealing process. Detailed fabrication procedures of the PDMS layer are provided in **Supplementary Note 3**. As illustrated in Fig. 3d, the PDMS gel forms a conformal and robust adhesion interface with the grating region. Upon pipeline deformation under external compression, the viscoelastic PDMS layer undergoes pronounced deformation, which amplifies the stress transfer efficiency to the LPFG region. Specifically, the viscoelastic stretching of the PDMS layer induces enhanced bending perturbations along the LPFG, thereby strengthening the modulation of the mode coupling condition and improving the stress-induced wavelength sensitivity. Meanwhile, the cured PDMS exhibits a combination of mechanical compliance and strong interfacial adhesion, ensuring synchronous deformation between the pipeline substrate and the fiber sensor. This effectively suppresses interfacial slippage and improves the linearity and repeatability of the resonance dip shift. Specifically, the viscoelastic stretching of the PDMS layer induces enhanced bending perturbations along the LPFG, thereby strengthening the modulation of the mode coupling condition and improving the stress-induced wavelength sensitivity [19]. Meanwhile, the cured PDMS exhibits a combination of mechanical compliance and strong interfacial adhesion, ensuring synchronous deformation between the pipeline substrate and the fiber sensor. This effectively suppresses interfacial slippage and improves the linearity and repeatability of the resonance dip shift.

As shown in **Fig. 3e**, the bending sensitivity of the LPFG resonance dip is significantly enhanced from −3.44 nm/cm to −8.97 nm/cm, with an improved linear fitting coefficient of 0.99, while maintaining an instantaneous response characteristic due to the intrinsic nature of optical signal transmission. In contrast, the bending-insensitive FBG exhibits negligible wavelength variation under bending perturbations (see **Fig. S4**), serving as a stable reference channel. This differential response significantly improves the bending detection limit (evaluated using $3\sigma$/sensitivity, where $\sigma = 0.02$ nm), reducing it from 0.017 cm to 0.006 cm. In addition, both LPFG and FBG exhibit good linear responses to temperature variations (**Fig. 3f**), while possessing distinct sensitivity coefficients (see **Fig. S5**). This differentiated spectral response provides the fundamental basis for constructing a multi-wavelength demodulation matrix, enabling accurate decoupling and simultaneous measurement of multiple physical parameters.

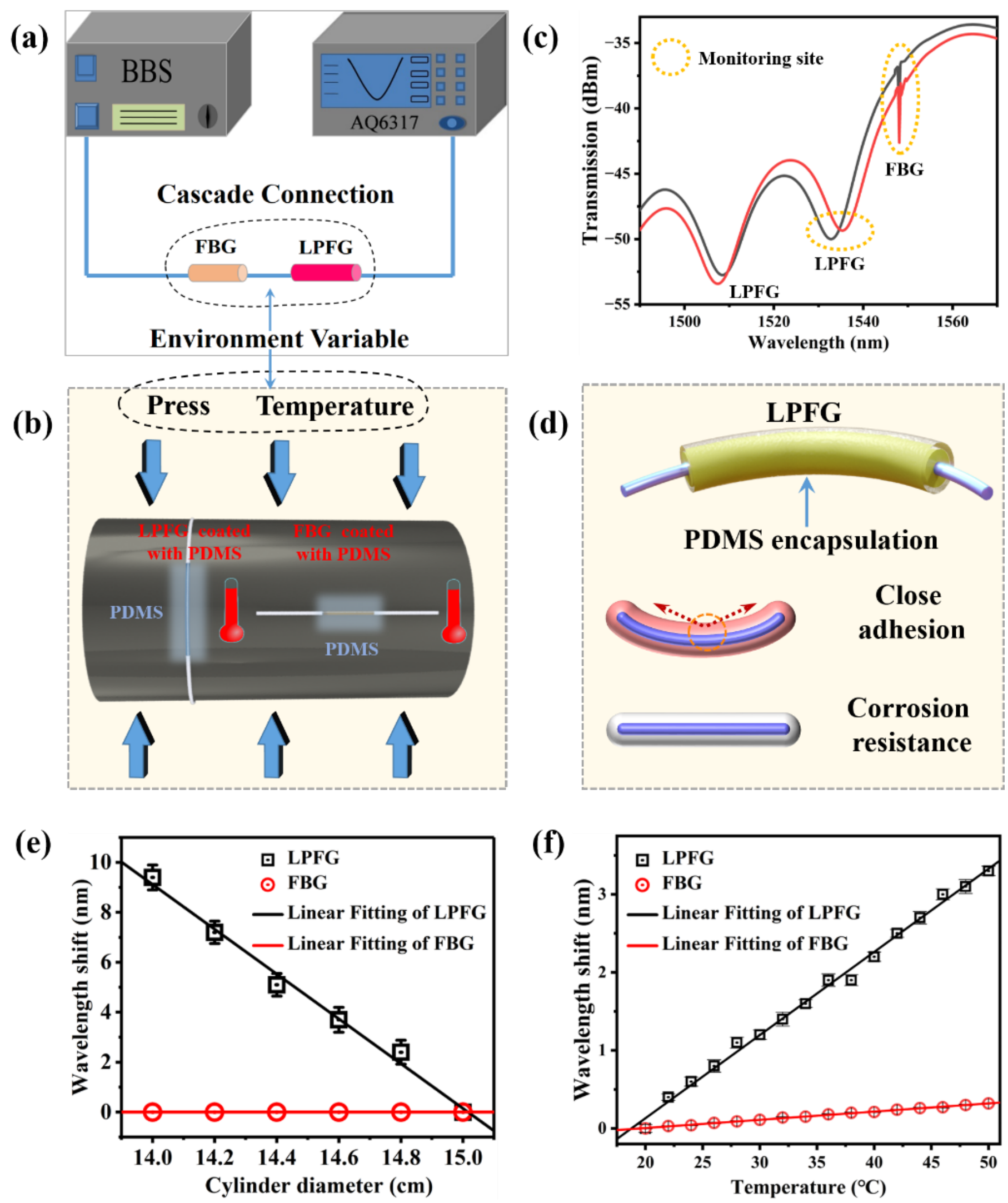

**Fig. 3** Temperature-compensated bending stress measurement system. (a) Schematic of the LPFG and FBG cascaded sensing system. (b) Transmission spectrum of the cascaded LPFG-FBG system. (c) Schematic of pipeline bending monitoring based on an LPFG-FBG cascaded sensing system. (d) Illustration of bending stress sensitivity enhancement using a polymer layer. (e) Temperature sensitivities of LPFG and FBG sensors. (f) Bending stress sensitivities of LPFG and FBG sensors.

Subsequently, a multi-wavelength matrix as shown in formula 10 is introduced for signal demodulation [20]. Cross-sensitivity is effectively overcome by exploiting spectrally distinct dips with differentiated sensitivities, enabling decoupled multi-parameter sensing via matrix inversion. The detailed demodulation procedure is provided in **Supplementary Note 6**.

$$\begin{bmatrix} \Delta\lambda_{\mathrm{Dip\ LPFG}} \\ \Delta\lambda_{\mathrm{Dip\ FBG}} \end{bmatrix} = \begin{bmatrix} \mathrm{K}_{\mathrm{T}_{\mathrm{Dip\ LPFG}}} & \mathrm{K}_{\mathrm{S}_{\mathrm{Dip\ LPFG}}} \\ \mathrm{K}_{\mathrm{T}_{\mathrm{Dip\ FBG}}} & \mathrm{K}_{\mathrm{S}_{\mathrm{Dip\ FBG}}} \end{bmatrix} \begin{bmatrix} \Delta \mathrm{T} \\ \Delta \mathrm{S} \end{bmatrix} \tag{10}$$

$\mathrm{K}_{\mathrm{T}_{\mathrm{Dip\ LPFG}}}$ is the sensitivity of LPFG dip to change with temperature, $\mathrm{K}_{\mathrm{S}_{\mathrm{Dip\ LPFG}}}$ is the sensitivity of LPFG dip to change with bending stress, and $\mathrm{K}_{\mathrm{T}_{\mathrm{Dip\ FBG}}}$ is the sensitivity of FBG dip to change with temperature. $K_{S_{Dip\ FBG}}$ is the sensitivity of FBG dip to change with bending stress. Substituting the corresponding parameters yields formula 11. The negative sign indicates a shift of the dip toward shorter wavelengths, while a positive sign denotes a shift toward longer wavelengths.

$$\begin{bmatrix} \Delta\lambda_{Dip\ LPFG} \\ \Delta\lambda_{Dip\ FBG} \end{bmatrix} = \begin{bmatrix} 0.106 & -8.971 \\ 0.011 & 0 \end{bmatrix} \begin{bmatrix} \Delta T \\ \Delta S \end{bmatrix} \tag{11}$$

The polymer-encapsulated cascaded grating system effectively suppresses temperature cross-sensitivity, enabling stable, high detection limit, real-time bending monitoring. Notably, the heterogeneous resonance dips induced by bending-assisted annealing in the LPFG provide a foundation for constructing a multi-parameter sensing matrix, allowing independent resolution of different physical perturbations. Furthermore, polymer encapsulation enhances mechanical coupling and expands sensing dimensionality, establishing a versatile platform for multi-parameter fiber-optic sensing. Based on this, a simultaneous sensing scheme integrating temperature, bending, and humidity is proposed. Here, bending reflects pipeline deformation under external pressure, while humidity serves as a critical factor associated with corrosion. The simultaneous monitoring of these parameters enables more comprehensive and accurate pipeline condition assessment.

**Scalable system for simultaneous multi-parameter monitoring**

As emphasized in the Introduction, the use of identical grating sensors is inherently unsuitable for multi-wavelength matrix sensing. Although multiple monitoring sites (e.g., peaks or dips) may exist, the intrinsic homogeneity of the same grating type leads to similar sensitivity responses to identical perturbations, preventing effective parameter decoupling. Achieving a high-quality multi-wavelength matrix therefore typically requires complex system architectures, which increases both demodulation difficulty and maintenance burden in practical applications. In this work, bending-assisted annealing introduces controlled heterogeneity between dual resonance dips of the LPFG, enabling differentiated sensitivity responses within a single grating. This strategy significantly simplifies the sensing architecture while maintaining high sensing performance, providing an efficient pathway toward compact multi-parameter sensing systems. **Fig. 4a** illustrates the integrated multi-parameter sensing configuration for simultaneous measurement of temperature, humidity, and pressure. In this system, the LPFG and FBG are strategically functionalized and spatially arranged to achieve selective sensitivity to different physical perturbations. Specifically, the LPFG region is encapsulated with PAAm, endowing it with humidity-responsive functionality, while the FBG is encapsulated with PDMS, ensuring mechanical stability and temperature sensitivity while remaining insensitive to bending [21]. The sensing architecture is designed such that temperature, humidity, and pressure induce distinct spectral responses across different resonance features. Temperature variations affect both LPFG and FBG due to intrinsic thermo-optic and thermal expansion effects. Humidity changes primarily modulate the RI and swelling behavior of the PAAm layer, selectively influencing the LPFG resonance dips (see **Supplementary Note 5** for details) [22]. Meanwhile, external pressure induces pipeline deformation, which is efficiently transferred to the LPFG region as bending perturbations, while the axially aligned FBG remains largely unaffected. This differentiated response mechanism establishes the basis for constructing a multi-wavelength sensing matrix, enabling simultaneous and decoupled extraction of temperature, humidity, and pressure signals from the measured spectral shifts.

Three characteristic spectral features, namely LPFG a, LPFG b, and the FBG resonance, were selected to monitor the responses to temperature, bending stress, and humidity variations, respectively **(Fig. 4b–g)**. The extracted wavelength shifts were incorporated into formula 12 to construct a three-wavelength demodulation matrix for multi-parameter sensing. Notably, LPFG a and LPFG b exhibit opposite wavelength shift directions and distinct sensitivity coefficients under external perturbations. This intrinsic heterogeneity between the dual resonance dips is critical for establishing independent sensing channels within a single LPFG. Combined with the fundamentally different sensing mechanism of the FBG, which serves as a temperature reference, the resulting multi-wavelength matrix is non-singular, enabling the simultaneous and decoupled measurement of temperature, bending stress, and humidity.

$$\begin{bmatrix} \Delta\lambda_{\mathrm{Dip\ LPFG\ a}} \\ \Delta\lambda_{\mathrm{Dip\ LPFG\ b}} \\ \Delta\lambda_{\mathrm{Dip\ FBG}} \end{bmatrix} = \begin{bmatrix} -0.106 & 0.097 & 0.011 \\ 7.821 & -8.732 & 0 \\ -0.405 & 0.788 & 0 \end{bmatrix} \begin{bmatrix} \Delta T \\ \Delta S \\ \Delta RH \end{bmatrix} \tag{12}$$

The responses to temperature and mechanical stress originate from the intrinsic properties of the grating structure, including the thermo-optic effect, thermal expansion, and photoelastic effect. Therefore, compared to the response time of the spectrometer, the fiber-optic sensor can be considered to exhibit a near-instantaneous response with negligible hysteresis. In contrast, the response speed of humidity sensing is governed by the adsorption and desorption dynamics of water molecules within the hygroscopic film. As shown in **Fig. S7**, when the relative humidity increases from 78% to 84%, the sensor reaches a stable response within 30 s, which is sufficient for long-term pipeline monitoring applications. Moreover, under constant humidity conditions, the wavelength drift remains limited to approximately 0.3 nm over 30 min or longer, demonstrating excellent stability and reliability for continuous operation.

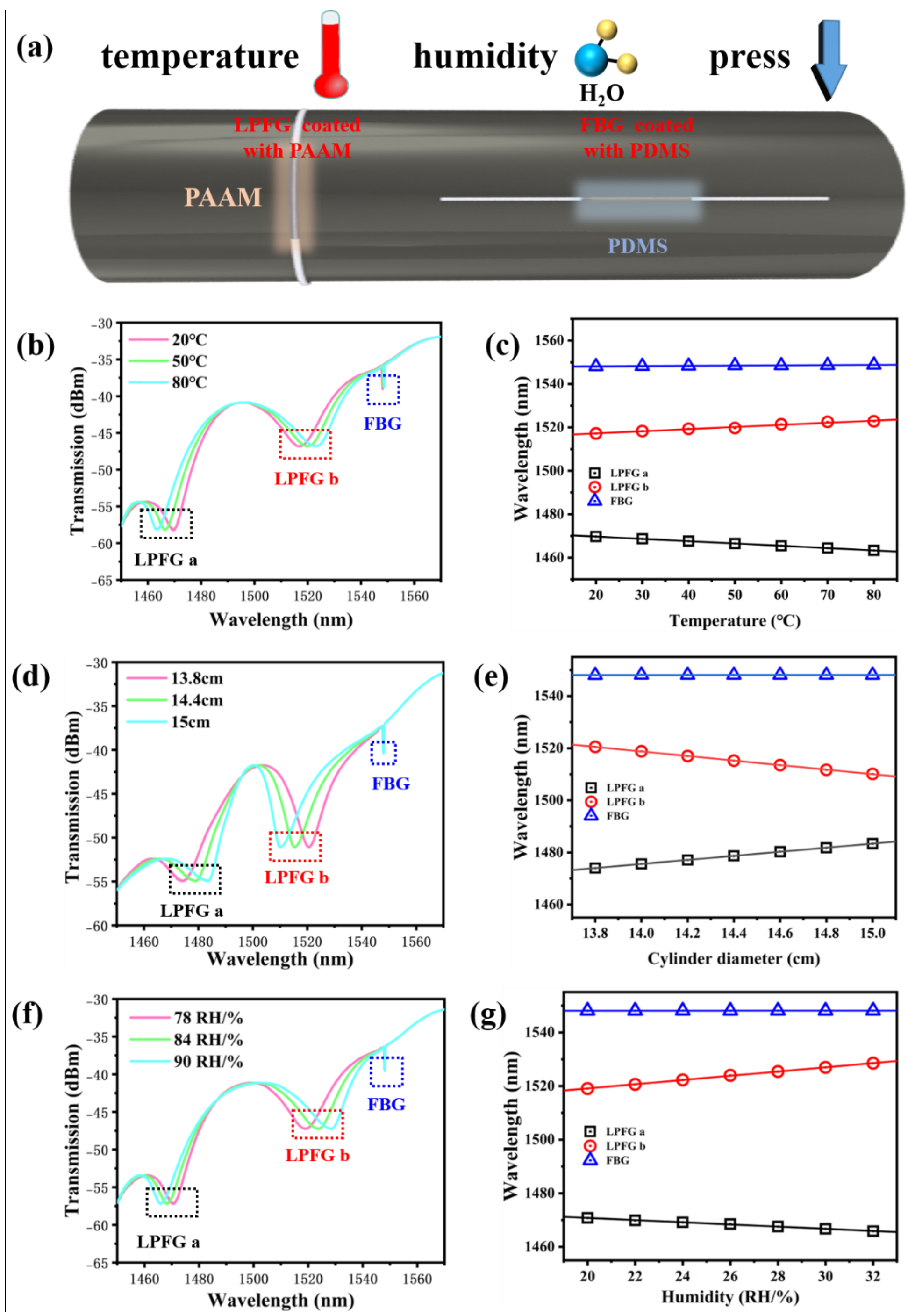

**Fig. 4** A simultaneous sensing system for bending, humidity, and temperature. (a) Schematic of the LPFG and FBG cascaded sensing system. (b) Cascaded system spectral variation with bending. (c) Three dips linear bending sensitivity relationship. (d) Cascaded system spectral variation with humidity. (e) Three dips linear humidity sensitivity relationship. (f) Cascaded system spectral variation with temperature. (g) Three dips linear temperature sensitivity relationship.

The proposed sensing system is not limited to pipeline monitoring applications. By tailoring surface-functionalized films and exploiting the intrinsic heterogeneity of LPFG sensing sites, the platform demonstrates strong potential for scalable multi-parameter sensing within a single device. This strategy effectively addresses the long-standing challenge of cross-sensitivity in fiber-optic sensors, while avoiding the complexity associated with conventional multi-sensor configurations and their corresponding demodulation difficulties. Overall, this work establishes a highly simplified yet robust paradigm for multi-parameter fiber-optic sensing, featuring excellent signal discrimination capability and enhanced system integration.

**Conclusion**

In this work, a dual-dip heterogeneous LPFG sensing system enabled by annealing under bending was developed, achieving simultaneous enhancement of sensitivity and intrinsic spectral heterogeneity. The bending sensitivity of the LPFG increased from −2.21 nm/cm to −3.47 nm/cm, while dual resonance dips exhibited opposite shift trends, providing the basis for multi-parameter demodulation.

Practical applications were demonstrated in both wearable and pipeline monitoring scenarios. For wrist sensing, a sensitivity of 7.1 nm/° was achieved, with a wavelength variation of only 0.3–0.5 nm at identical bending states. In pipeline monitoring, sensitivities of −3.44 nm/cm and −3.38 nm/cm were obtained, with linearity coefficients of 0.98 and 0.96, respectively. By integrating PDMS encapsulation and a cascaded LPFG–FBG structure, the bending sensitivity was further enhanced to −8.97 nm/cm with a linearity of 0.99, and the detection limit was improved from 0.017 cm to 0.006 cm. The FBG served as a stable temperature reference, enabling effective temperature compensation. Furthermore, PAAm functionalization enabled simultaneous sensing of temperature, bending, and humidity. A non-singular three-wavelength matrix was constructed based on the heterogeneous sensitivities of three spectral dips (LPFG a, LPFG b, and FBG). The humidity response stabilized within 30 s (78%–84% RH), with wavelength drift below 0.3 nm over 30 min, demonstrating fast response and high stability.

Overall, this work establishes a compact and scalable multi-parameter fiber sensing paradigm, effectively mitigating cross-sensitivity while simplifying system complexity, with strong potential for structural health monitoring and harsh-environment sensing.

**Acknowledgements**

The authors acknowledge the support by the National Natural Science Foundation of P. R. China (12104092) and the Fujian Medical University Talent Startup Fund (XRCZX2019032).

## Supplementary Materials

# A Dual-Dip Heterogeneous LPFG Sensing System via Annealing under Bending with Temperature and Humidity Compensation

Cuiying Huang[1], Riming Xu[2], Jialing Kang[3], Weihan Chen[3], Xingnan Chen[4], Yanbo Li[5], Jin Wang[2,6]*

1. School of Arts and Sciences, Fujian Medical University, Fuzhou 350122, China.
2. Physical Sciences and Engineering Division, King Abdullah University of Science and Technology, Thuwal 23955-6900, Saudi Arabia.
3. School of Basic Medical Sciences, Fujian Medical University, Fuzhou 350122, China.
4. Collage of Mechanical and Electrical Engineering, Hainan Vocational University of Science and Technology, Haikou 571125, China.
5. Department of Civil and System Engineering, Johns Hopkins University, Baltimore 21218, USA.
6. College of Physics and Optoelectronics, Faculty of Science, Beijing University of Technology, Beijing 100124, China.

*Corresponding author: Jin Wang
E-mail: wangjinbjut321@163.com

## Supplementary Note 1: Materials and film fabrication

Polydimethylsiloxane (PDMS) base and curing agent were mixed at a 10:1 weight ratio and stirred thoroughly until a homogeneous mixture was obtained. The mixture was then degassed under vacuum to remove trapped air bubbles. The degassed PDMS was subsequently applied to encapsulate the device region and cured at elevated temperature to form an elastomeric protective layer.

The hydrogel precursor was prepared using deionized water, acrylamide (AAm), poly (ethylene glycol) diacrylate (PEGDA), and ammonium persulfate (APS). Specifically, 2 g of acrylamide was dissolved in 5 g of deionized water, followed by the addition of 10 μL PEGDA as an auxiliary crosslinking component. The mixture was stirred thoroughly at room temperature until a homogeneous transparent solution was obtained. Subsequently, 10 mg APS was added as the initiator and dissolved completely under continuous stirring. The resulting precursor solution was then immediately used for in situ polymerization. During device fabrication, the precursor solution was applied onto the grating region and polymerized under elevated temperature simultaneously with the grating annealing process. Under the synchronized annealing condition, the precursor underwent thermally initiated polymerization and gradually formed a hydrogel network, thereby enabling in situ solidification and integration of the hydrogel on the grating surface.

Fiber Bragg grating (FBG) was fabricated with a Bragg wavelength centered at approximately 0.5 μm. The protective polymer coating was retained in the grating region to preserve mechanical robustness and environmental stability.

Long-period fiber grating (LPFG) was fabricated with a grating period of approximately 500 μm and a duty cycle of 1:1. To facilitate direct interaction with the surrounding environment, the protective polymer coating was selectively removed in the grating region. The LPFG was attached to the wall of a cylindrical tube and annealed at a constant temperature of 200 °C. This bending annealing process promotes more effective stress relaxation, enabling the $SiO_2$ matrix to release the intrinsic residual stresses introduced during the fiber drawing process. In this study, a cylinder with a curvature radius of 15 cm was selected for the bending annealing. Annealing under excessively small curvature radii may result in grating damage or instability in mode coupling. After annealing, the stress–strain behavior of the fiber becomes more linear and the structural stability is improved, which is advantageous for fiber-optic sensors requiring long-term stability and linear response characteristics.

When polymer encapsulation is employed, the polymer is first prepared according to the aforementioned proportions and then coated onto the surface of the grating sensor according to the requirements of the intended application. The polymer heating process can be performed simultaneously and in situ with the grating annealing process, as illustrated in Fig. S1.

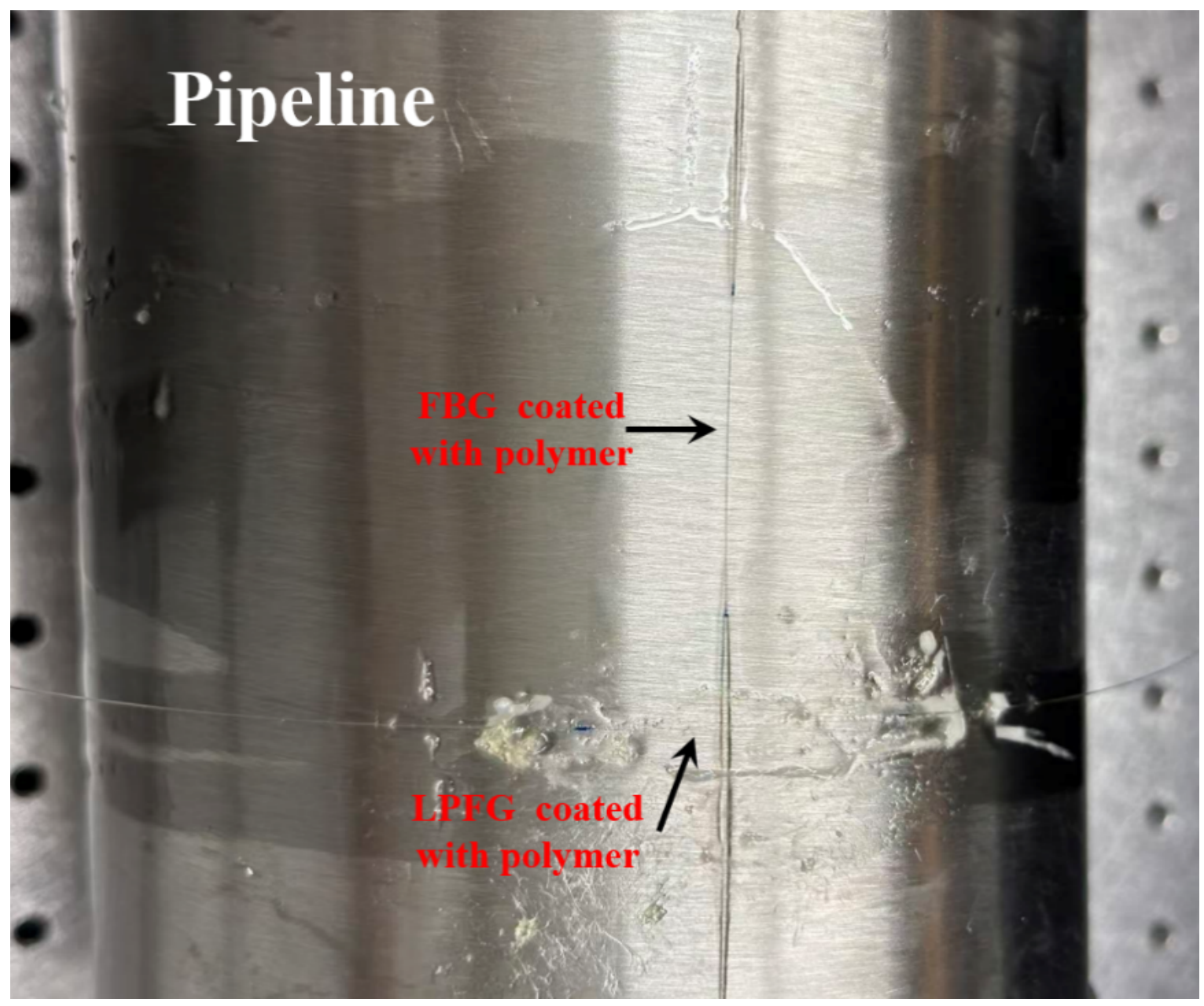

Supplementary Fig. 1 Cascaded gratings encapsulated with polymer film on a pipeline.

## Supplementary Note 2: Measurement setup

A broadband light source (1400–1700 nm) was employed as the input, and the transmitted spectra were acquired using a spectrometer (AQ6317C) with a wavelength range of 600–1750 nm and a resolution of 0.02 nm, ensuring precise spectral characterization. Sensor sensitivities under each condition were obtained by averaging multiple repeated measurements. The schematic diagram of the experimental setup is shown in Fig. S2.

For temperature characterization, a temperature-controlled chamber was employed to regulate the ambient temperature. The fiber-optic sensing system was securely mounted on thermally stable holders to maintain the grating region flat and prevent deformation.

For humidity measurements, the fiber-optic sensing system was fixed on fiber holders and placed inside a humidity-controlled chamber. The humidity sensor and the fiber sensing region were co-located to ensure measurement consistency.

For bending sensing, the optical fiber sensor was mounted on a compressible and stretchable cylindrical surface, with the LPFG attached along the radial direction and the FBG aligned along the axial direction.

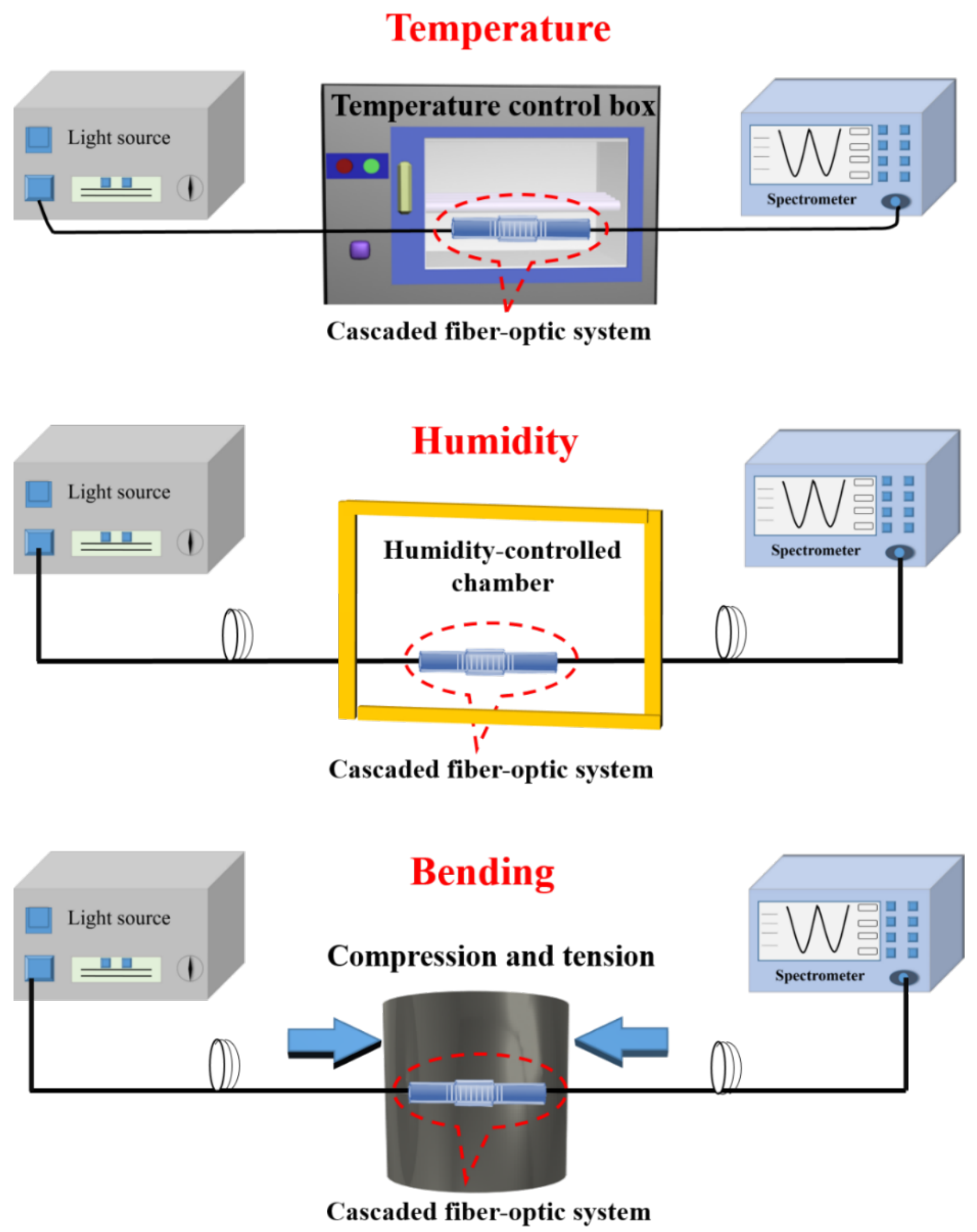

Supplementary Fig. 2 Schematic of the fiber-optic sensing setup for temperature, humidity, and bending measurements.

**Supplementary Note 3: SEM morphology and EDS elemental analysis of the PDMS film**

The microstructure and elemental composition of the PDMS encapsulation layer were examined by scanning electron microscopy (SEM) and energy-dispersive X-ray spectroscopy (EDS) in Fig. S3. The SEM image reveals a featureless, compact, and uniform morphology, indicative of a continuous and defect-free polymer coating on the substrate. Elemental mapping confirms the homogeneous distribution of C, O, and Si throughout the film, with no detectable phase segregation or compositional inhomogeneity, underscoring the high structural integrity of the PDMS layer.

This structurally uniform coating establishes a robust and conformal mechanical interface between the optical fiber and the host substrate, enabling efficient transduction of external stimuli. The intrinsic viscoelasticity and strong interfacial adhesion of PDMS further promote enhanced microbending-induced perturbations, thereby reinforcing stress–optical coupling and contributing to the improved sensing performance, in agreement with the mechanism discussed in the main text.

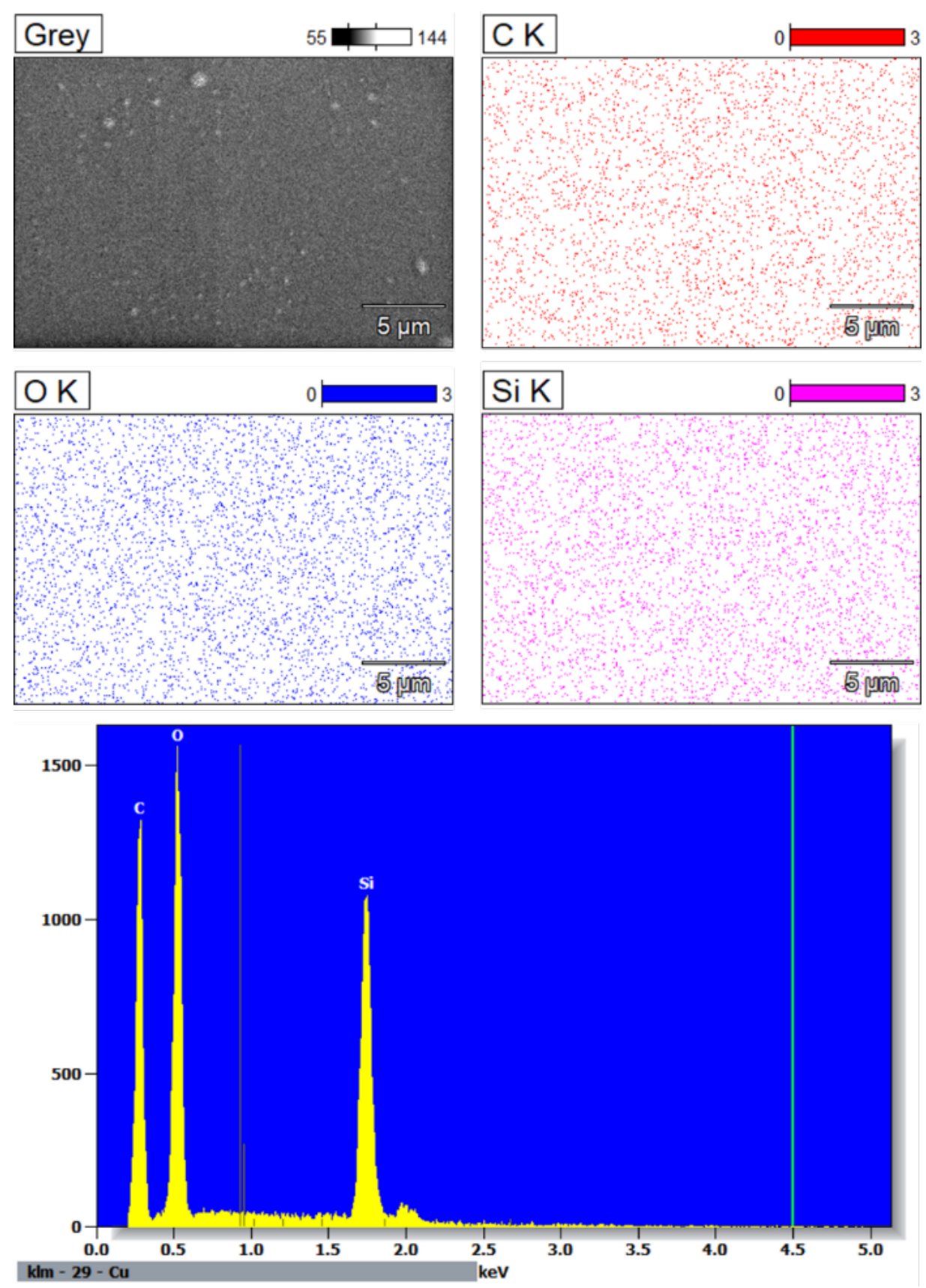

Supplementary Fig. 3 SEM morphology and EDS elemental analysis of the PDMS thin film.

**Table S1 EDS-derived elemental composition of PDMS**

| Element | Counts | Weight % | Atom % |
|---|---|---|---|
| C | 19570 | 40.84 | 47.07 |
| N | 1523 | 14.10 | 13.94 |
| O | 6308 | 45.06 | 38.99 |
| Total | | 100.00 | 100.00 |

## Supplementary Note 4: Temperature and stress sensitivity measurements of LPFG and FBG

Fig. S4 presents the microbending sensitivity characterization of the LPFG and FBG. As the cylinder diameter decreases, the LPFG transmission spectrum exhibits a pronounced redshift of the resonance dip, indicating a strong response to microbending perturbations. The extracted wavelength shift shows a clear linear dependence on the cylinder diameter, yielding a sensitivity of −8.97 nm/cm with a high linear fitting coefficient of 0.99.

In contrast, the FBG demonstrates negligible spectral variation under identical microbending conditions. Only minor fluctuations are observed in the resonance region, which can be attributed to spectrometer noise rather than structural perturbations. This distinct difference in response confirms that the LPFG serves as a highly sensitive microbending sensing element, while the FBG functions as a mechanically insensitive reference channel, thereby enabling improved signal-to-noise ratio in microbending detection. All sensitivity coefficients were calibrated under single-variable variation, for instance with the stress maintained constant during temperature changes.

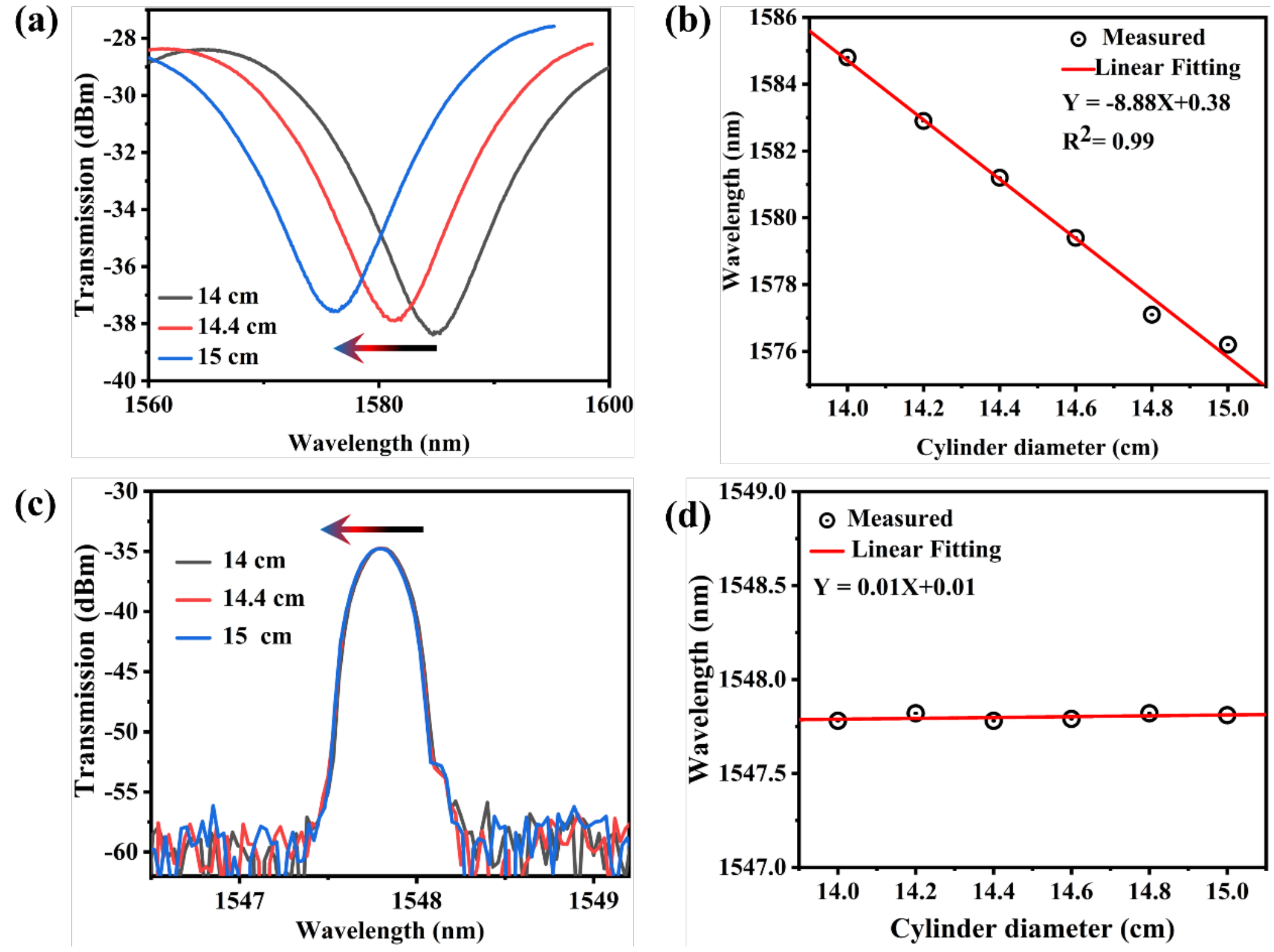

Supplementary Fig. 4 Micro bending sensitivity measurement of LPFG and FBG. (a) LPFG spectral variation with micro bending. (b) Trough of LPFG linear micro bending sensitivity relationship. (c) FBG spectral response remains insensitive to micro bending. (d) Only minor disturbances are observed at the FBG trough, mainly attributed to the spectrometer noise.

Fig. S5 presents the temperature-dependent spectral responses of the LPFG and FBG. As the temperature increases from 20 to 50 °C, both the LPFG resonance dip and the FBG Bragg wavelength exhibit a gradual redshift, as shown in Fig. S4a and S4c, respectively. This behavior originates from the combined effects of the thermo-optic coefficient and thermal expansion of the fiber material.

The wavelength shifts of both sensors show a clear linear dependence on temperature. As illustrated in Fig. S4b, the LPFG exhibits a temperature sensitivity of ~0.106 nm/°C with a high linear fitting coefficient ($R^2$ = 0.99). Similarly, the FBG response in Fig. S4d demonstrates a linear temperature sensitivity of ~0.011 nm/°C, also with $R^2$ = 0.99.

Notably, the LPFG shows a significantly higher temperature sensitivity compared to the FBG, reflecting their distinct sensing mechanisms. This difference in sensitivity coefficients provides the necessary basis for temperature compensation and multi-parameter decoupling in the cascaded sensing system. All sensitivity coefficients were calibrated under single-variable variation, for instance with the temperature maintained constant during stress changes.

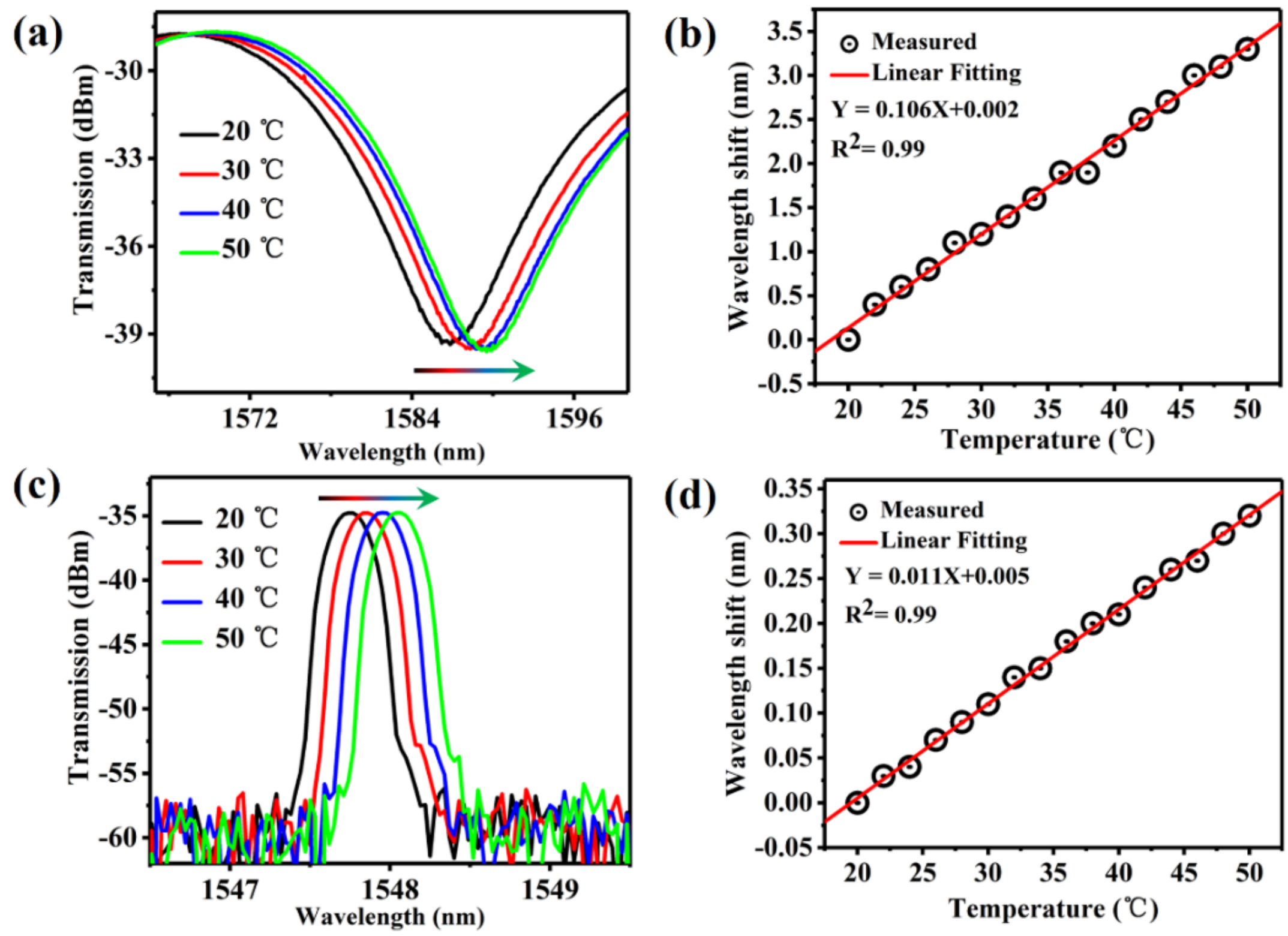

Supplementary Fig. 5 Temperature sensitivity measurement of LPFG and FBG**.** (a) LPFG spectral variation with temperature. (b) Trough of LPFG linear temperature sensitivity relationship. (c) FBG spectral variation with temperature. (d) Trough of FBG linear temperature sensitivity relationship.

**Supplementary Note 5: SEM morphology and EDS elemental analysis of the PAAM film**

The microstructure and elemental composition of the PAAm film were characterized by SEM and EDS in Fig. S6. The SEM image reveals a uniform and featureless surface morphology, indicating the formation of a continuous and homogeneous coating. Elemental mapping shows a homogeneous distribution of C, N, and O elements across the film, with no observable aggregation or phase separation, suggesting good compositional uniformity. The corresponding EDS spectrum exhibits characteristic peaks consistent with the chemical composition of PAAm, confirming the successful formation of the polymer layer.

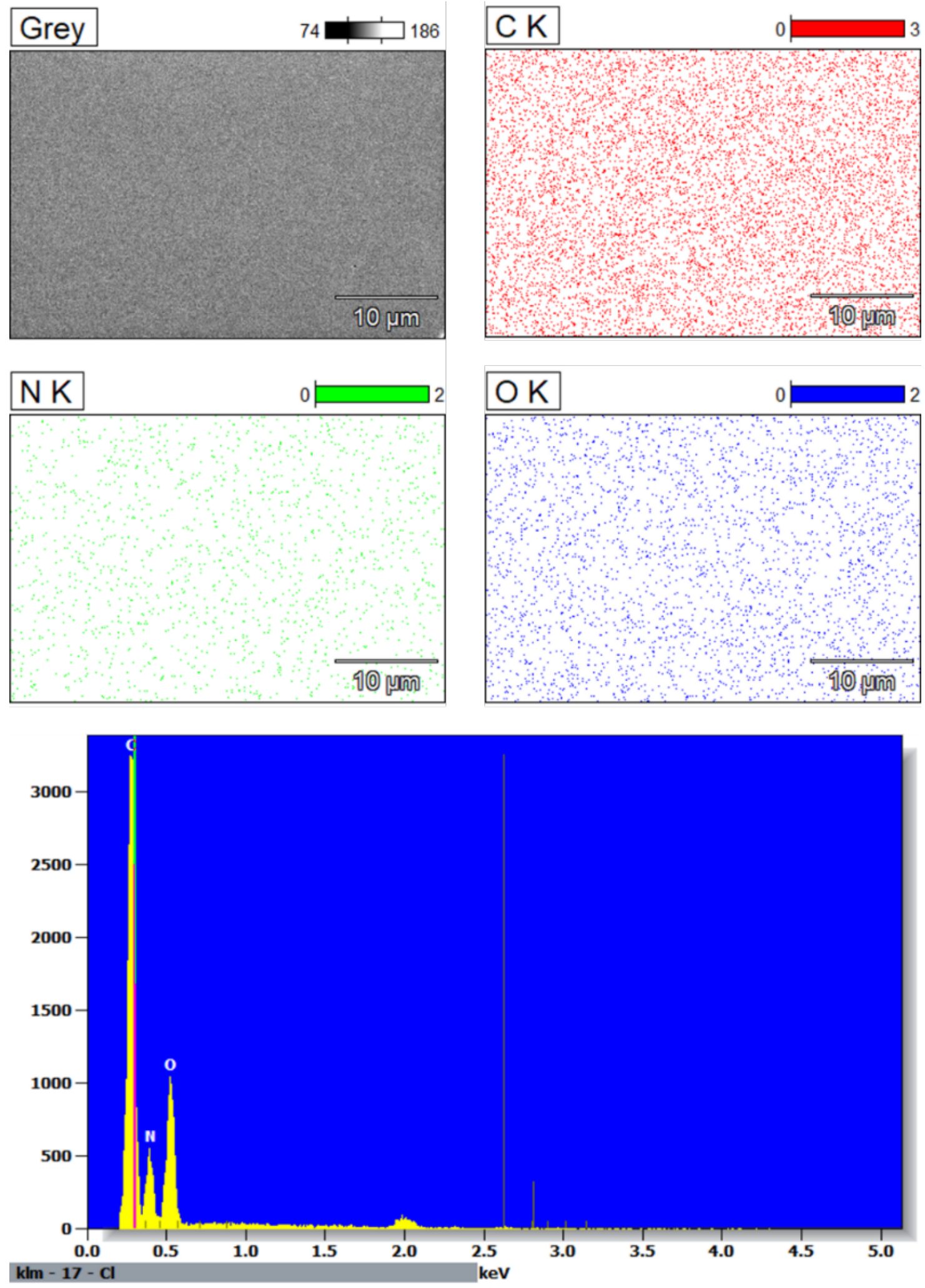

Supplementary Fig. 6 SEM morphology and EDS elemental analysis of the PAAM thin film.

**Table S2 EDS-derived elemental composition of PAAM**

| Element | Counts | Weight % | Atom % |
|---|---|---|---|
| C | 7531 | 18.24 | 28.51 |
| O | 9097 | 33.34 | 39.12 |
| Si | 9432 | 48.42 | 32.37 |
| Total | | 100.00 | 100.00 |

**Supplementary Note 6: Response time and stability of humidity sensing.**

Fig. S7 illustrates the dynamic response and long-term stability of the sensor under varying humidity conditions. When the relative humidity (RH) increases from 78% to 84%, the resonance dip exhibits a clear redshift within 30 s, indicating a rapid response to humidity variation. Such a response speed is sufficient to meet the requirements for real-time and continuous pipeline monitoring. To further evaluate stability, the spectral responses were recorded at constant humidity levels over extended durations. At both 78% RH and 84% RH, the resonance wavelength shows only a slight drift of approximately 0.3 nm over 30 minutes or longer, demonstrating excellent temporal stability. This minimal fluctuation suggests that the sensor maintains a stable equilibrium state after initial response, with negligible long-term degradation or hysteresis.

Overall, the combination of fast response and high stability confirms the suitability of the sensor for long-term humidity monitoring applications, particularly in environments where reliable and continuous detection is required.

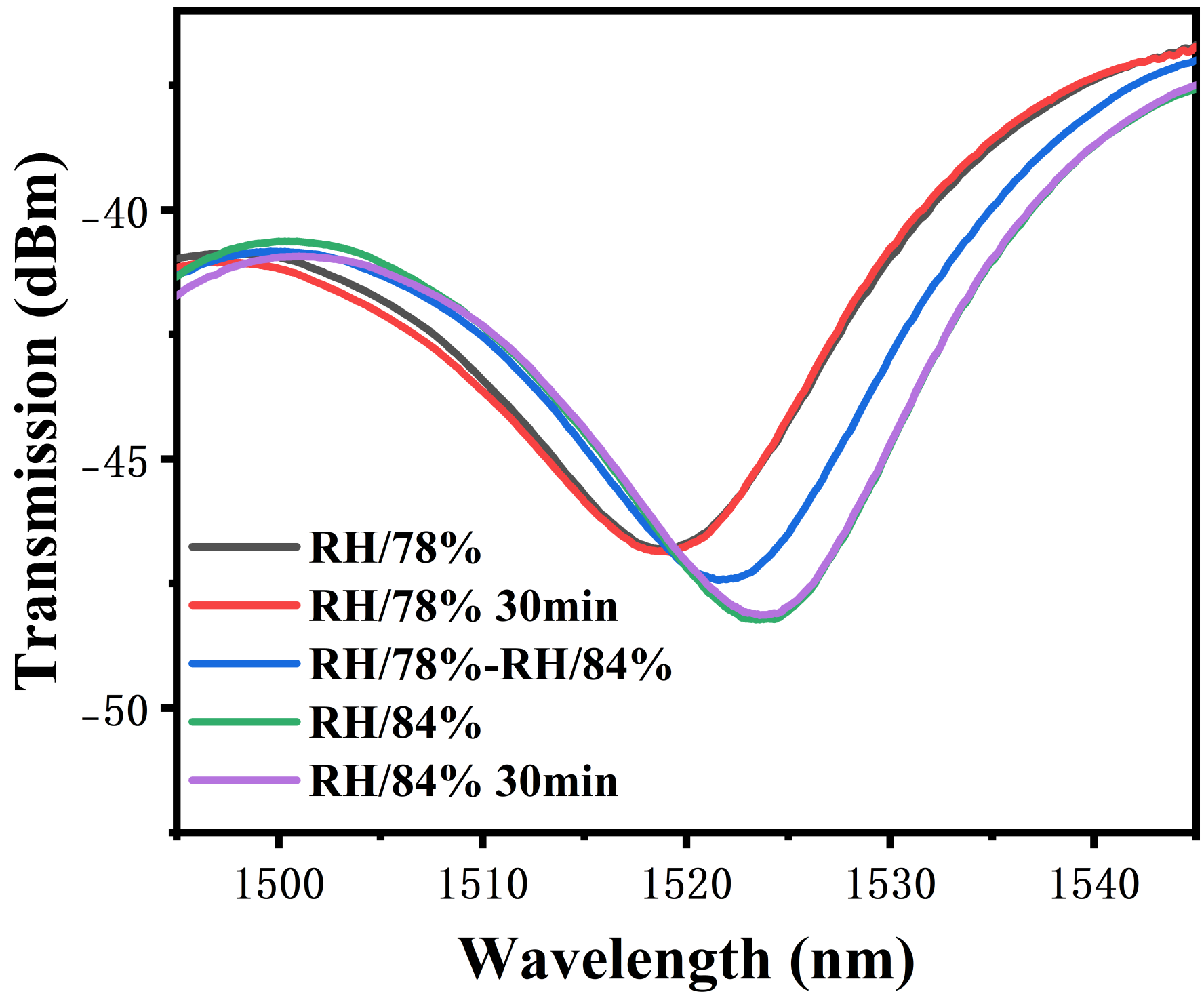

Supplementary Fig. 7 Stability and response time of humidity measurement.

## Supplementary Note 7: Demodulation mechanism of the multi-wavelength matrix

Simultaneous variations in temperature, microbending, and humidity cannot be resolved from a single spectral dip due to intrinsic cross-sensitivity in fiber-optic sensing. Here, instead of assigning individual dips to specific parameters, three spectrally distinct dips with differentiated sensitivities are employed for multi-parameter sensing, whose sensitivity disparity ensures the construction of a non-singular matrix. Following calibration, a 3 × 3 sensitivity matrix is established, and the environmental variations are retrieved by inverting the matrix using the measured wavelength shifts. This multi-wavelength matrix approach enables accurate and decoupled determination of temperature, microbending, and humidity. It should be noted that bending-assisted annealing may exacerbate the cross-coupling between temperature and stress in the LPFG sensing system. Therefore, once the sensing system is fixed, calibration must be performed prior to operation before applying the multi-wavelength matrix method.